\documentclass[useAMS,usenatbib]{mn2e}
\usepackage{natbib}
\usepackage{graphicx}
\usepackage{amsmath}
\usepackage[T1]{fontenc}
\DeclareGraphicsExtensions{.pdf}

\title[21 cm emission - LAE cross-power spectrum]{Predictions for the 21cm-galaxy cross-power spectrum observable with LOFAR and Subaru}
\author[D. Vrbanec et al.]{Dijana Vrbanec$^{1}$,\thanks{E-mail:
dvrbanec@mpa-garching.mpg.de} Benedetta Ciardi$^{1}$, Vibor Jeli\'{c}$^{2,3,4}$, Hannes Jensen$^{5}$, \newauthor Saleem Zaroubi$^{2}$,  Elizabeth R. Fernandez$^{2}$, Abhik Ghosh$^{2}$, Ilian T. Iliev$^{6}$, \newauthor Koki Kakiichi$^{1}$, L\'{e}on V. E. Koopmans$^{2}$, Garrelt Mellema$^{5}$
\\
$^{1}$Max Planck Institute for Astrophysics, Karl-Schwarzschild-Strasse 1, D-85748 Garching bei M\"{u}nchen, Germany\\
$^{2}$Kapteyn Astronomical Institute, University of Groningen, PO Box 800, NL-9700 AV Groningen, the Netherlands\\
$^{3}$ASTRON, PO Box 2, NL-7990 AA Dwingeloo, the Netherlands\\
$^{4}$Ru{\dj}er Bo\v{s}kovi\'{c} Institute, Bijeni\v{c}ka cesta 54, 10000 Zagreb, Croatia \\
$^{5}$Depatment of Astronomy and Oskar Klein Centre, Stockholm University, AlbaNova, SE-10691 Stockholm, Sweden\\
$^{6}$Astronomy Centre, Department of Physics and Astronomy, Pevensey II Building, Univesity of Sussex, Falmer, Brighton BNI 9QH}
\begin{document}

\date{Accepted - Received -; in original form -}

\pagerange{\pageref{firstpage}--\pageref{lastpage}} \pubyear{2014}

\maketitle

\label{firstpage}

\begin{abstract}
The 21cm-galaxy cross-power spectrum is expected to be one of the promising probes of the Epoch of Reionization (EoR), as it could offer information about the progress of reionization and the typical scale of ionized regions at different redshifts. With upcoming observations of 21cm emission from the EoR with the Low Frequency Array (LOFAR), and of high redshift Ly$\alpha$ emitters (LAEs) with Subaru's Hyper Suprime Cam (HSC), we investigate the observability of such cross-power spectrum with these two instruments, which are both planning to observe the ELAIS-N1 field at $z=6.6$. In this paper we use N-body + radiative transfer (both for continuum and Ly$\alpha$ photons) simulations at redshift 6.68, 7.06 and 7.3 to compute the 3D theoretical 21cm-galaxy cross-power spectrum, as well as to predict the 2D 21cm-galaxy cross-power spectrum expected to be observed by LOFAR and HSC. Once noise and projection effects are accounted for, our predictions of the 21cm-galaxy cross-power spectrum show clear anti-correlation on scales larger than $\sim 60 h^{-1}$~Mpc (corresponding to $k\sim 0.1 h$~Mpc$^{-1}$), with levels of significance $p=0.04$ at $z=6.6$ and $p=0.048$ at $z=7.3$. On smaller scales, instead, the signal is completely contaminated.

\end{abstract}

\begin{keywords}
galaxies: high redshift - cosmology:observations - reionization - intergalactic medium
\end{keywords}

\section{Introduction}

The Epoch of Reionization (EoR) is one of the greatest observational frontiers in modern astrophysics. It corresponds to the transition from a neutral to an ionized Universe, as mostly young,  star-forming galaxies reionized the intergalactic medium (IGM) surrounding them. 

The early stages of cosmic reionization are characterized by the presence of isolated HII regions, which grow in number and size as structure formation progresses. During the latest stages of the process, the ionized regions are very large, with individual sizes exceeding tens of comoving Mpc. Eventually, they overlap and IGM reionization is complete (e.g. \citealt{Gnedin1997, Ciardi2000, Gnedin2000, Ciardi.Ferrara.White_2003, Maselli2003, Santos2004, Furlanetto2004, Haiman2005,  Iliev2006, McQuinn2007, Zahn2011, Ciardi.Bolton.Maselli.Graziani_2012, Iliev2014}). 
While early simulations were only a few Mpc in size \citep{Gnedin1997, Ciardi2000, Gnedin2000}, novel codes for cosmological N-body and hydrodynamical simulations and for radiative transfer finally enable reionization simulations with volumes larger than $\sim$ 100~Mpc \citep{Iliev2006, McQuinn2007, Iliev2014}, allowing for a correct abundance of rare massive haloes \citep{Barkana2004, Li2007, Trac2011}, while resolving also dwarf-size galaxies with masses $\sim 10^8$~M$_{\odot}$, which are considered to be the main sources of ionizing photons \citep{Volonteri2009, Loeb2009, Robertson2010, Fontanot2012}. In recent years, a significant improvement has taken place in terms of dynamic range of such simulations and sophistication in the implementation of various relevant physical processes (e.g. \citealt{Trac2011}).

Absorption spectra of high-redshift quasars suggest that reionization was completed by $z\approx 6$ \citep{Fan2006, Bolton2011, McGreer2015}. On the other hand, measurements of the primordial Cosmic Microwave Background (CMB) radiation obtained by the WMAP satellite indicate that the process started much earlier, suggesting that the Universe was neutral until $z=10.1\pm1.0$, if instantaneous reionization is assumed \citep{Komatsu2011}. More recent measurements with Planck instead give $z=8.8\pm1.1$ \citep{Planck2015}. Present observations though do not offer much information neither on the progress of reionization nor on the main sources responsible for it. Detection of the 21cm line from neutral hydrogen promises to offer insight in this respect.                      

Because of the wealth of information such observations could provide, 
there are significant efforts to detect  reionization by mapping the 21cm hyperfine transition line of neutral hydrogen with radio arrays such as 
LOFAR\footnote{http://www.lofar.org} \citep{Haarlem2013}, MWA\footnote{http://web.haystack.mit.edu/arrays/MWA}, PAPER\footnote{http://eor.berkeley.edu}, GMRT\footnote{http://gmrt.ncra.tifr.res.in} and SKA\footnote{http://www.skatelescope.org}. Calculations predict that the cosmological 21cm signal from the EoR will be extremely faint, while the system noise and the foregrounds will be orders of magnitude larger (e.g. \citealt{Madau1997, Ciardi2003, Zaldarriaga2004, Furlanetto2006a, McQuinn2006, Mellema2006, Morales2006, Jelic2008, Geil2008, Labropoulos2009, Harker2010, Geil2011, Chapman2012, Zaroubi2012, Bernardi2009, Pober2013, Parsons2014}). Due to the low signal to noise ratio (which for LOFAR is $\sim$ 0.2, \citealt{Labropoulos2009, Zaroubi2012}) the first observations of the 21cm signal will measure only statistical properties, such as the rms and power spectrum of the brightness temperature and their evolution with time \citep{Ciardi2003, Morales2004, Barkana2005, Bowman2006, McQuinn2006, Pritchard2007, Jelic2008, Pritchard2008, Harker2009, Harker2010}. Other probes of reionization that will be possible with observations of the redshifted 21cm line are imaging (e.g. \citealt{Koopmans2015, Mellema2015, Wyithe2015}), which will not be possible before SKA, 21cm forest \citep{Mack2012,Ciardi2013, Ciardi2015}, and cross-correlation with other observations at different wavelengths, such as near-infrared background radiation (NIRB; e.g. \citealt{Fernandez2014, Mao2014}), kinetic Sunyaev-Zel'dovich effect (kSZ; e.g. \citealt{Jelic2010, Tashiro2010}), galaxies (e.g. \citealt{Lidz2009, W2013, Park2014}), CO line (e.g. \citealt{Righi2008, Visbal2010, Lidz2011}) and CII line (e.g. \citealt{Gong2012, Silva2015}). Detection of such cross-correlations can provide further insight into different aspects of the EoR, such as the progress of reionization and the redshift at which the process is halfway, the evolution of the neutral hydrogen content, and the typical scale of ionized regions at different redshifts.

Another way to explore reionization is to probe high-$z$, young, star-forming galaxies, which are considered to be the dominant sources of ionizing photons. Such galaxies are expected to have a strong Ly$\alpha$ emission line due to the interaction of the interstellar medium with ionizing radiation from young massive stars \citep{Partridge1967}. Depending on the detection method, such galaxies are typically referred to as Ly$\alpha$ emitters (LAEs) and Lyman break galaxies (LBGs). Star-forming galaxies that are luminous enough to be detected with existing telescopes most likely populate fairly massive dark matter haloes, with masses in excess of $10^{10}$~M$_{\odot}$ \citep{Dijkstra2014}. They ionize their surroundings forming large HII bubbles in which one or more star-forming galaxies reside (e.g. \citealt{Dijkstra2014}). Ly$\alpha$ photons emitted by those galaxies can therefore propagate and redshift away from line resonance through the ionized IGM before entering the neutral IGM \citep{Miralda1998, Santos2004, Cen2000, Barton2004, Cen2005, McQuinn2007, Iliev2008, Mesinger2008, Curtis-Lake2012}. These photons are then less likely to be scattered out of the line of sight. This is why LAE luminosity functions \citep{Haiman1999, Malhotra2004, Haiman2005, LeDelliou2005, Furlanetto2006, LeDelliou2006, Mao2007, Dijkstra2007, Jensen2013, Jensen2014}, number density \citep{Malhotra2006}  and clustering \citep{Furlanetto2006, McQuinn2007, Wyithe2007, Jensen2013} are the main methods to study the EoR with galaxies. A reduction in the number of observed sources, and thus a suppression of the luminosity function, is expected with increasing redshift, due to the larger amount of neutral gas in the IGM (e.g. \citealt{Haiman1999}).

As of now, 207 LAEs have been observed at $z$=6.45-6.65 \citep{Ouchi2010}, 1 at z=6.96 \citep{Ota2008} and 7 at $z$=7.3 \citep{Konno2014}, while there are no confirmed LAEs at higher redshifts. These numbers are likely to increase soon, as the Subaru telescope with its new Hyper Suprime-Cam (HSC) started narrow-band observations at redshift 6.6 and 7.3. The HSC has a 1.5 deg (in diameter) Field of View (FoV) and it will observe LAEs in its Deep ($z$=6.6, for a total of  28 deg$^2$) and Ultra-deep ($z$=6.6 and 7.3, for a total of 3.4 deg$^2$ at each redshift) fields \citep{Miyazaki2012}. It is estimated that HSC will observe $\sim$ 5500 LAEs at $z$=6.6 and $\sim$ 40 LAEs at $z$=7.3 (Ouchi 2012, private communication).

Although detection of the 21cm signal and observations of LAEs will provide invaluable insight on reionization and its sources, the shape and normalization of their cross-power spectrum can offer additional information \citep{Lidz2009, W2013, Park2014}\footnote{\citet{W2013} have shown that LAEs are more promising probes of reionization than LBGs.}. 
In this article we will present theoretical 3D 21cm-galaxy cross-power spectra computed from full radiative transfer + N body simulations by \citet{Iliev2014}, post-processed with a Ly$\alpha$ radiative transfer code by \citet{Jensen2013} and \citet{Laursen2010}. Differently from previous papers, here we accurately model LAEs, so that their simulated properties reproduce the observed ones.
We will also show 2D 21cm-galaxy cross-power spectra from mock observations obtained by adding instrumental effects for the LOFAR and Subaru telescopes. The paper is organized as follows: in section 2 we describe the simulations used; in section 3 we show the theoretical 3D cross-power spectra as well as the observed 2D cross-power spectra; in section 4 we show 21cm-galaxy cross-correlation functions. In section 5 we discuss the noise from both LOFAR and HSC observations, and draw conclusions in section 6. 

The following set of cosmological parameters was used: $\Omega_{\Lambda}=0.73, \Omega_m= 0.27, \Omega_b=0.044, h=0.7$ and $\sigma_8=0.8,n_s=0.96$, consistent with WMAP 5-year data \citep{Komatsu2009}.

\section{Simulations}

To compute the 21cm-galaxy cross-power spectrum we have used a full radiative transfer + N-body simulation of reionization \citep{Iliev2014} in a box of comoving length $425 h^{-1}$~Mpc (corresponding to $\sim$ 4 deg at $z$=7) with 165 billion particles distributed on a grid of $10,976^3$ cells ($3.9\,h^{-1}$kpc gravity force resolution) and a radiative transfer grid of $504^3$ cells.
The N-body simulation has been run from redshift $z=300$ to $z=2.6$, with initial conditions generated using the Zel'dovich approximation and a power spectrum of the linear fluctuations given by the CAMB code \citep{Lewis2000}. This simulation was then used as an input to the radiative transfer code C$^2$-RAY \citep{Mellema2006a} to follow the reionization history of the IGM. More specifically, the halo catalogues were used to construct the sources of ionizing radiation as in \citet{Iliev2007}. As the minimum resolved halo mass is $M_{h,min}= 10^9$~M$_{\odot}$, haloes with masses of $10^8-10^9$ M$_{\odot}$ were modeled as a sub-grid population \citep{Iliev2014, Ahn2015}. All haloes were assigned an ionizing photon emission rate per unit time, $\dot{N}_{\gamma}$, proportional to the halo mass, M$_h$:
\begin{equation}
 \dot{N}_{\gamma}=\frac{g_{\gamma}M_h\Omega_b}{\Omega_0 m_p}\left(\frac{10\ \mathrm{Myr}}{\Delta t}\right),
\end{equation}
where $m_p$ is the proton mass, $\Omega_b$ and $\Omega_0$ have their usual cosmological meaning, $\Delta t=11.46$ Myr is the time between two snapshots of the N-body simulation, and $g_{\gamma}$ is a source efficiency coefficient that incorporates the star formation efficiency, the total photon production per stellar baryon and the ionizing photon escape fraction \citep{Iliev2006,Jensen2014}. Haloes with masses down to $10^9$ M$_{\odot}$ were assigned a source efficiency of $g_{\gamma}=1.7$. Smaller sources with masses down to $10^8$ M$_{\odot}$ were assigned $g_{\gamma}=7.1$, to account for a lower metallicity and a more top-heavy initial mass function, but they were assumed to be suppressed within ionized regions (for ionization fraction higher than 10 per cent; \citealt{Iliev2014}). The radiation emitted by the sources is propagated through the gridded density field, and the distribution of neutral hydrogen is obtained at various redshifts. This is used to calculate the associated differential brightness temperature according to the usual formalism (e.g. \citealt{Field1959,Madau1997,Furlanetto2006a}): 
\begin{multline}
\delta T_b= 28.5\ \mathrm{mK}\ (1+\delta)x_{HI}\left(\frac{\Omega_b}{0.042}\frac{h}{0.73}\right)\\\times\left[\left(\frac{1+z}{10}\right)\left( \frac{0.24}{\Omega_m} \right)\right]^{1/2},
\end{multline}
where $x_{HI}(1+\delta)=n_{HI}/\langle n_H\rangle$ is the mean density of neutral hydrogen in units of the mean density of hydrogen at redshift $z$. 

For our purposes, we used boxes from the simulation at $z=$6.68, 7.06 and 7.3, corresponding to volume (mass) averaged ionized fractions $\langle x \rangle=$ 0.93 (0.95), 0.65 (0.73) and 0.48 (0.58), respectively. These particular boxes were chosen because HSC will have two narrow-band filters observing at redshifts 6.6 and 7.3, while 7.06 is an intermediate value. 

The same simulations were processed with a Ly$\alpha$ radiative transfer code to model high-$z$ LAEs and study their observability. Motivated by detailed radiative transfer calculations by \citet{Laursen2011}, the Ly$\alpha$ line was modeled as a double peaked profile with little emission at the line centre, and a width that depends on the halo mass. Intrinsic luminosities were calibrated against observations, with a model where the Ly$\alpha$ luminosities of haloes of a given mass follow a log-normal distribution with a mean that is proportional to the halo mass. 
After assigning an intrinsic Ly$\alpha$ spectrum to the dark matter haloes in the N-body simulations, the observed luminosities are calculated including the attenuation from the IGM along a large number of lines of sight from each of the haloes (for more details we refer the reader to the original papers \citealt{Jensen2013, Jensen2014}). From the same work we extracted the Ly$\alpha$ intrinsic and transmitted luminosities, which we use to produce HSC mock observations.

In the computation of the 3D cross-power spectra we merged bins to obtain $\Delta k>0.02\ h$~Mpc $^{-1}$, which corresponds to the smallest mode resolved by a FoV of 16~deg$^2$, i.e. equivalent to our simulations. We also made sure to avoid correlations in power due to the window function\footnote{The sphere used to compute a spherically averaged $P(k)$ in a simulation of comoving length 425~$h^{-1}$Mpc, must be equivalent in volume and thus have a radius $R=264$ $h^{-1}$Mpc comoving. A window function for a spherical tophat has its first zero at $dk\cdot R\sim4.5$, so that $k$-values spaced by less than $4.5/R=0.02\ h$~Mpc$^{-1}$ will be correlated \citep{Feldman1994, Furlanetto2007,Lidz2009, W2013}.} by using a binning with $\Delta \log k=0.02$.
As the FoV used to compute the 2D cross-power spectra is smaller (i.e. 7~deg$^2$ and 1.7~deg$^2$ at $z=$6.6 and 7.3, respectively; see section 5), we used $\Delta \log k=0.03$ (0.05) and  $\Delta k>0.04$ (0.07) $h$~Mpc $^{-1}$ for $z=6.6$ (7.3).

\section{Cross-power spectrum}

In this section we present our calculations of the theoretical and observational cross-power spectra.

At each redshift, the 21cm-galaxy cross-power spectrum at wave number $k=|k|$, $\Delta^2_{21,gal}(k)$, can be decomposed into three contributing terms (e.g. \citealt{Lidz2009}):
\begin{multline}
\Delta^2_{21,gal}(k)=\tilde{\Delta}^2_{21,gal}(k)/\delta T_{b0} \\  = \langle x_{HI} \rangle [ \Delta^2_{x_{HI},gal}(k)+\Delta^2_{\rho,gal}(k)+\Delta^2_{x_{HI}\rho,gal}(k)],
\end{multline}
where $\Delta^2_{x_{HI},gal}$, $\Delta^2_{\rho,gal}$ and $\Delta^2_{x_{HI}\rho, gal}$ are the neutral fraction-galaxy, density-galaxy and neutral density-galaxy cross-power spectra, respectively. $\delta T_{b0}$ is the 21cm brightness temperature relative to the CMB for a fully neutral gas element at the mean cosmic density, and $\langle x_{HI} \rangle$ is the volume-averaged neutral hydrogen fraction. $\Delta^2_{a,b}$ is the dimensionless cross-power spectrum of two random fields, $a$ and $b$, and it is equal to $\Delta^2_{a,b}(k)=k^3P_{a,b}(k)/{2\pi^2}$ for the 3D cross-power spectrum, and $\Delta^2_{a,b}(k)=2\pi k^2P_{a,b}(k)$ for the 2D power spectrum. $P_{a,b}$ represents the dimensional cross-power spectrum between fields $a$ and $b$. The latter are represented in terms of their fractional fluctuations at a location $r$, i.e. $\delta_a(r)=(a(r)-\langle a\rangle)/\langle a \rangle$, and similarly for $b$\footnote{Note that we evaluate the theoretical cross-power spectrum with $\langle
a \rangle=(\sum_{i=1}^N a_i)/N$, where $N$ is the number of pixels in the portion of the simulation used. All the quantities are calculated like this, with the exception of the galaxy field in mock observations, which is instead calculated using $\langle N_{gal} \rangle =  N_{gal}/V$, where $N_{gal}$ is the number of galaxies in the mock observation, and $V$ is the volume of the survey. This was done for an easier comparison with the shot noise power spectrum $P_{shot}(k)=1/n_{gal}$, where $n_{gal}$ is the average number of galaxies in the survey volume.}. A more detailed discussion of the various terms can be found in \citet{Lidz2009}. 

\subsection{Theoretical 21cm-galaxy cross-power spectrum}

\begin{figure}
\includegraphics[width=84mm]{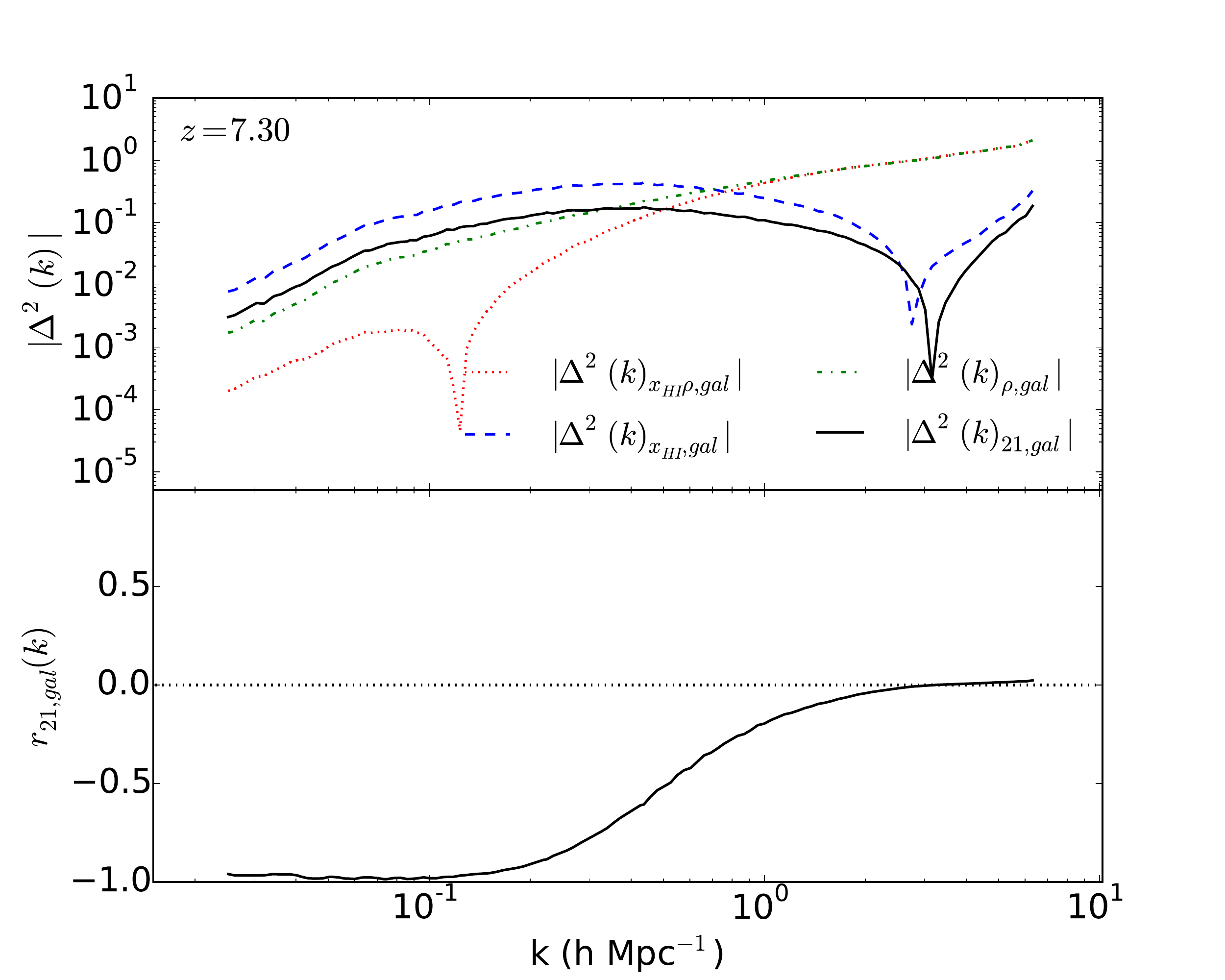}
\caption{{\bf Top panel:} spherically averaged 3D 21cm-galaxy cross-power spectrum, $\Delta^2_{21,gal}$ (black solid line) at $z=7.3$, together with its contributing terms, i.e. the neutral hydrogen-galaxy cross-power spectrum, $\Delta^2_{x_{HI},gal}$ (blue dashed), the density-galaxy cross-power spectrum, $\Delta^2_{\rho , gal}$ (green dashed-dotted) and the neutral density-galaxy cross-power spectrum, $\Delta^2_{x_{HI}\rho ,gal}$ (red dotted). {\bf Bottom panel:} 21cm-galaxy cross-correlation coefficient, $r_{21,gal}$ (black solid), and  zero-correlation coefficient (black dotted).}
\label{21t0.58}
\end{figure}

To understand the 21cm-galaxy cross-power spectrum we first show the theoretical spherically averaged 3D 21cm-galaxy cross-power spectrum at $z=7.3$ (which was computed using haloes with M$_h > 10^{10}$ M$_{\odot}$, i.e. 3 million galaxies at $z=6.68$, 2.3 million at $z=7.06$ and 1.9 million at $z=7.3$), together with its contributing terms (Fig. \ref{21t0.58}, top panel) and the corresponding cross-correlation coefficient (Fig. \ref{21t0.58}, bottom panel), defined as $r_{21,gal}(k)=P_{21,gal}(k)/[P_{21}(k)P_{gal}(k)]^{1/2}$. This corresponds to the ideal case in which all galaxies could be observed.
From the behaviour of $\Delta^2_{\rho, gal}$ it is clear that, as expected, the galaxies are strongly correlated with the density field on small scales, because galaxy formation is biased toward high density regions, while the correlation decreases as we move towards larger scales, but always remains positive. The neutral hydrogen-galaxy cross-power spectrum, $\Delta^2_{x_{HI},gal}$, instead, is negative on large scales where there is a paucity of galaxies but most of the HI resides. A turn around is observed in correspondence of the typical scale of the HII regions, and then the correlation drops off since the hydrogen inside such regions is completely ionized independently from the number of sources. $\Delta^2_{x_{HI}\rho,gal}$ is positive on the largest scales and becomes negative towards smaller scales, where it cancels out with $\Delta^2_{\rho, gal}$.  
The final 21cm-galaxy cross-power spectrum thus follows the shape of $\Delta^2_{x_{HI},gal}$ on small scales, and that of $\Delta^2_{x_{HI},gal}$ and $\Delta^2_{\rho, gal}$ on large scales. 
The cross-correlation coefficient (bottom panel of Fig. \ref{21t0.58}) shows more clearly that the 21cm signal and the high-$z$ galaxies are anti-correlated on large scales, and become uncorrelated on scales smaller then the typical size of the ionized regions. 

Similar conclusions were drawn by \citet{Lidz2009} and \citet{W2013}, although our results are closer to those of \citet{Lidz2009} because of the lower resolution employed in the simulation by \citet{W2013}. 

\begin{figure}
\includegraphics[width=84mm]{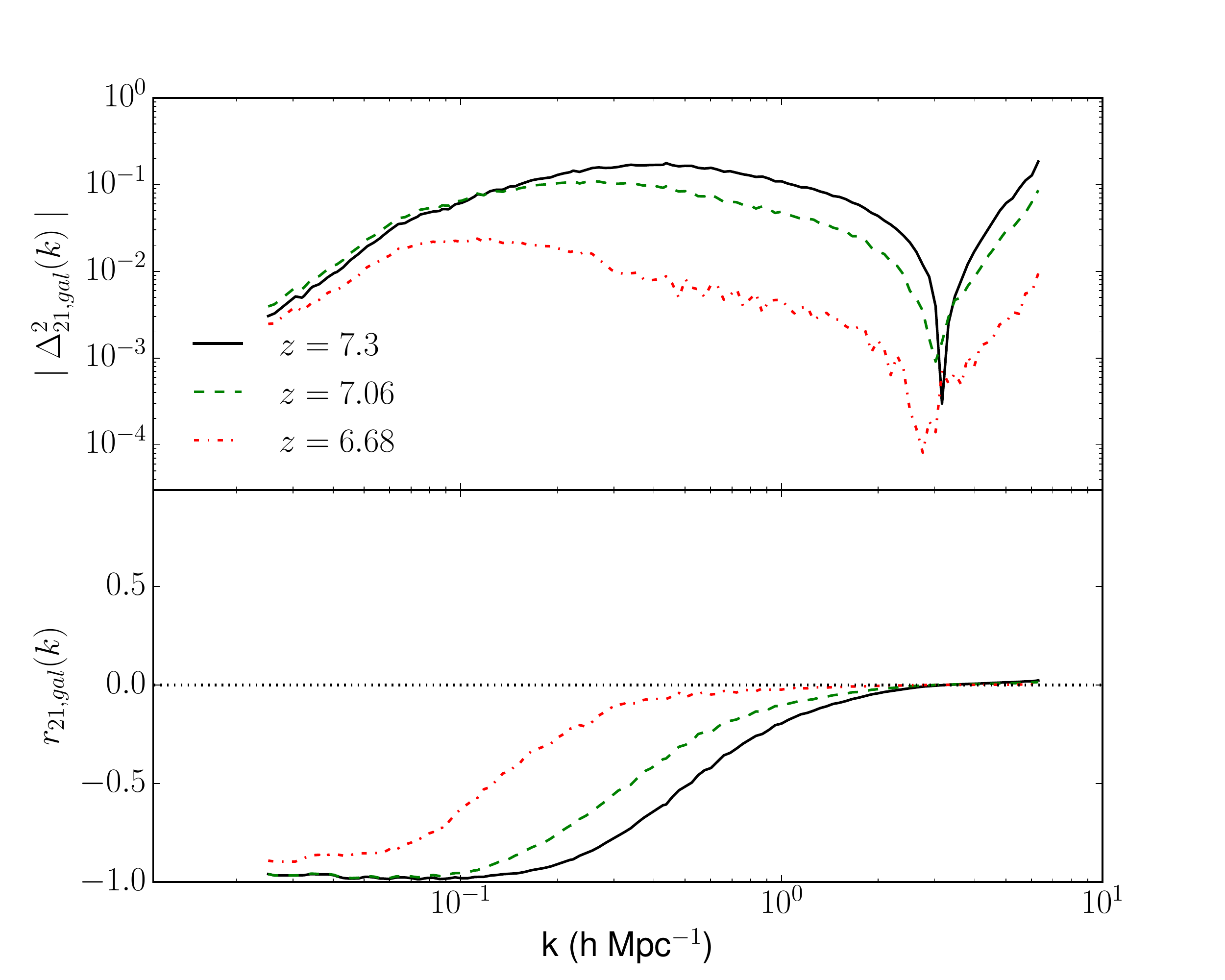}
\caption{{\bf Top panel:} spherically averaged 3D 21cm-galaxy cross-power spectrum, $\Delta^2_{21,gal}$,  at $z=7.3$ (black solid line), 7.06 (green dashed), and 6.68 (red dashed-dotted). {\bf Bottom panel:} 21cm-galaxy cross-correlation coefficient, $r_{21,gal}$, corresponding to $\Delta^2_{21,gal}$. }
\label{21tall}
\end{figure}

Figure \ref{21tall} shows $\Delta^2_{21,gal}$ and $r_{21,gal}$ for the chosen redshifts. We can see that the amplitude of the power spectrum decreases with decreasing redshift, while the turnover point shifts towards larger scales. This indicates that, as reionization proceeds, the anti-correlation decreases because of the paucity of neutral hydrogen, and the ionized bubbles grow in size. This is more clearly seen in the behaviour of the cross-correlation coefficients, which shift towards smaller $k$ with decreasing redshift.

\subsection{Observed 21cm-LAE cross-power spectrum}

In this section we will show our predictions for the 2D 21cm-LAE cross-power spectrum as it would be observed with LOFAR and HSC. To do that, we added projection effects and constrained the galaxy number density to match HSC expectations, and we added noise to the 21cm field to simulate LOFAR observations.

HSC will probe the reionization epoch with the Ultra-deep and Deep layers of the HSC Survey. Observations are made with narrow-band filters ($\Delta z=0.1$, equivalent to approximately one tenth of our simulation length {$\approx 42 \ h^{-1} $Mpc), so that the LAEs redshift will be tightly constrained. Because the LAEs detected with a particular filter will be observed as if they were lying on a single plane, the observed 21cm-LAE cross-power spectrum will be a circularly averaged 2D cross-power spectrum. HSC will observe 4 fields of 7 deg$^2$ at redshift $z=6.6$ as part of a Deep layer, and 4 fields of 1.7 deg$^2$ (two at $z=6.6$ and two at $z=7.3$) as part of a Ultra-deep layer (Ouchi 2012, private communication). One of the fields in the Deep layer is ELAIS-N1, which will also be observed with LOFAR \citep{Jelic2014}. 

We reduced the box dimension to match the HSC's field of view (7 deg$^2$ at $z=6.6$ and 1.7 deg$^2$ at $z=7.3$) by removing external cells\footnote{The choice of removing external cells is arbitrary and we have checked that it does not affect the final results.}. We then divided our simulation boxes of brightness temperature and galaxies into 10 sub-boxes of 50 slices each, corresponding to a $\Delta z=0.1$. Each sub-box obtained from the galaxy simulation is collapsed onto a single plane to mimic the fact that HSC observations will provide a 2D map of galaxies. This map is then correlated with each of the 50 slices of the corresponding brightness temperature sub-box to obtain 50 2D 21cm-galaxy cross-power spectra, which are then averaged to mimic the result of observations of a single FoV. From the 10 sub-boxes we then obtain 10 2D 21cm-galaxy cross-power spectra, which can again be averaged so that our results are not sample dependent. 

Figure \ref{2Dtheo} shows final, unnormalized by $\delta T_{b0}$, 2D 21cm-galaxy cross-power spectra before including the noise and the constraints on the galaxy number density\footnote{We note that while the solid lines represent the absolute value of the average cross-power spectrum (i.e. the average could be both positive and negative), the shaded area is obtained from the scatter in absolute averaged values (i.e. only positive numbers). For this reason the solid lines do not always lie at the center of the shaded areas.}. Even in 2D the cross-power spectra still retain much of their shape, although some features are lost due to projection effects and reduction in field of view, e.g. the turnover point is not clear anymore. Projection effects also induce a reduction in the value of the anti-correlation, clearly observed in the cross-correlation coefficient, which drops from $r_{21,gal}\approx-1$ to $r_{21,gal}\approx-0.5$. 

\begin{figure}
\includegraphics[width=84mm]{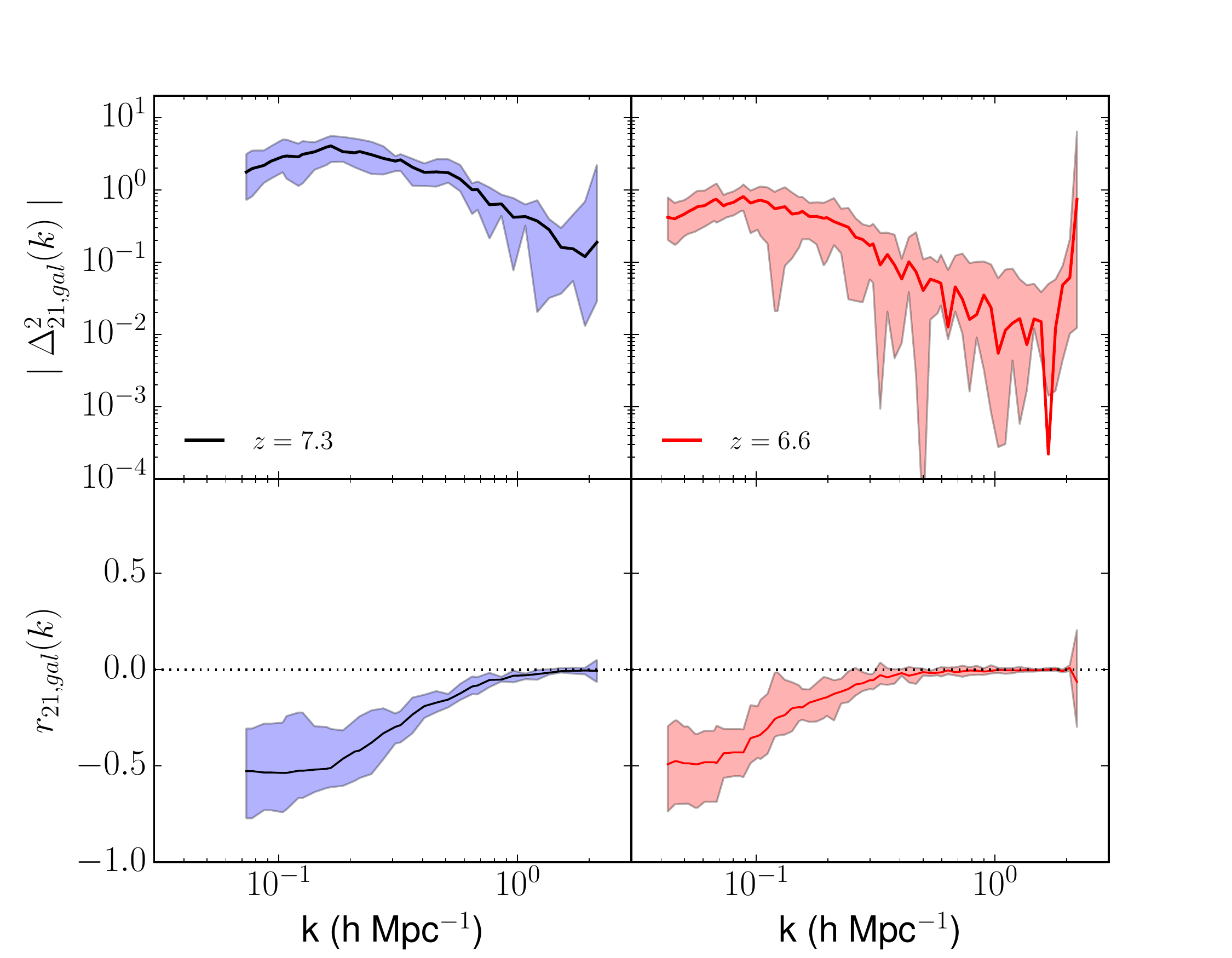}
\caption{{\bf Top panels:} 
2D unnormalized, circularly averaged 21cm-galaxy cross-power spectra at $z=7.3$ (left panel) and 6.6 (right). Shaded areas indicate scatter from 10 mock observations. {\bf Bottom panels:} 21cm-galaxy cross-correlation coefficient, $r_{21,gal}$, corresponding to $\Delta^2_{21,gal}$.
}
\label{2Dtheo}
\end{figure}

When selecting LAEs for our mock observations we assign intrinsic equivalent widths (EW) to the LAE sample according to a log-normal distribution, as was done by \citet{Jensen2014}. The distribution is designed to approximately fit observations made by \citet{Jiang2013}, while giving 65 \% of the galaxies EW below 25~\r{A} (consistent with \citealt{Stark2010}, as shown in Fig. 1 in \citealt{Jensen2014}). We first selected all the galaxies with EW $>20$ \r{A} (consistent with HSC expectations), and among these only the 1375 (20) most luminous ones at $z=6.6$ (7.3), to match the number expected to be observed by HSC.

LOFAR will be detecting the cosmological 21cm signal with a field of view of 5 $\times$ 5 deg$^2$, and an angular resolution of 3.5$'$ \citep{Zaroubi2012}. To simulate the LOFAR noise at each frequency, we filled a LOFAR measurement set (the real and imaginary parts of the visibilities) with Gaussian random numbers. This was then imaged (Fourier transformed, accounting for the proper weighting) to obtain noise maps in real-space, and their root mean square was normalised according to (e.g. \citealt{Taylor1999}):
\begin{equation}
\sigma_n=\frac{W}{\eta_s}\frac{\mathrm{SEFD}}{\sqrt{2N(N-1)\Delta\nu t_{int}}},
\label{eq:noise}
\end{equation}
where $W$ is a factor used to increase the noise according to the adopted weighting scheme, $\eta_s$ is the system efficiency, SEFD is the system equivalent flux density, $N$ is the number of stations, $\Delta \nu$ is the bandwidth, and $t_{int}$ is the integration time. Based on empirical SEFD values for LOFAR (e.g. $3000~{\rm Jy}$ at $150~{\rm MHz}$ towards the zenith; \citealt{Haarlem2013}), we expect $\sigma_n$ to be about $76~{\rm mK}$ at a resolution of $3.5~{\rm arcmin}$, at $150~{\rm MHz}$, after $600$ hours and $0.5~{\rm MHz}$ of integration and assuming $N=48$, $W=1.3$, $\eta_s=0.9$. Note that adopted noise values are indicative only, and they may change in the actual observations due to e.g. time-variable station projection losses of sensitivity, smaller system efficiency, etc. (\citealt{Haarlem2013}). More details about simulating the LOFAR noise can be found in e.g. \citet{Patil2014}. The simulated LOFAR noise was added to the brightness temperature map from the simulation.

In Figure \ref{21gal_lofar} we plot the resulting 2D unnormalized, circularly averaged 21cm-LAE cross-power spectra with (solid lines) and without (dashed) LOFAR noise. Despite the spectra being much noisier than the previous ones at all scales, a dependence of the normalization on redshift (i.e. amount of HI) and an anti-correlation ($r_{21,gal}\approx-0.20$) are still visible on large scales, with levels of significance of $p=0.04$ at $z=6.6$ and $p=0.048$ at $z=7.3$, although the turnover point can not be clearly identified. Even without LOFAR noise, observations at small scales will still be largely affected by shot noise and will not offer any reliable data (see sec.~\ref{sec:discuss}). From this analysis we conclude that only scales larger than $\sim 60$ (45) $h^{-1}$ Mpc, i.e. $k< 0.1$ (0.14) $h$~Mpc$^{-1}$, at $z=$6.6 (7.3) can be used for cross-correlation studies.

\begin{figure}
\includegraphics[width=84mm]{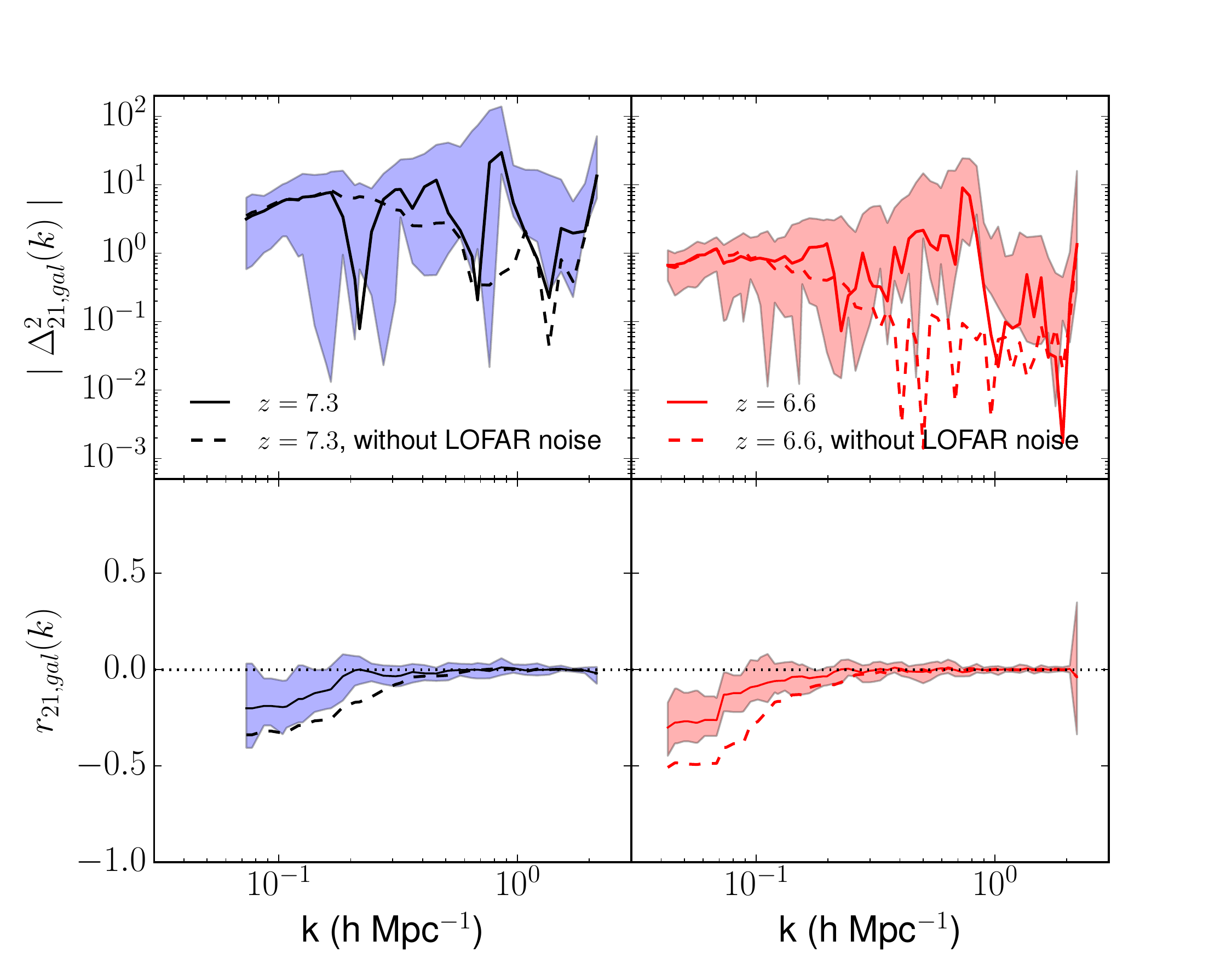}
\caption{{\bf Top panels:} 2D unnormalized, circularly averaged 21cm-LAE cross-power spectra at $z=7.3$ (left panel) and 6.6 (right). Shaded areas indicate scatter from 10 mock observations. Solid (dashed) lines refer to the cross-power spectra with (without) LOFAR noise. {\bf Bottom panels:} 21cm-galaxy cross-correlation coefficient, $r_{21,gal}$, corresponding to $\Delta^2_{21,gal}$. }
\label{21gal_lofar}
\end{figure}

For a HSC FoV equal to the one of LOFAR, though, we would expect the detection of 3140 and 90 LAEs at $z=6.6$ and 7.3, respectively. In this case (see Fig.~\ref{wishful}) the overall noise would be reduced, the anti-correlation signal would be stronger ($r_{21,gal}\approx-0.30$), large scales could be more reliably used, and information could be extracted down to $\sim 60$ (30) $h^{-1}$ Mpc, i.e. $k> 0.1$ (0.2) $h$~Mpc$^{-1}$, at $z=6.6$ (7.3). In addition, also information at scales larger than $\sim 130$ (80) $h^{-1}$ Mpc at $z=6.6$ (7.3) and up to $\sim 310 \ h^{-1}$ Mpc   would be available.
 
\begin{figure}
\includegraphics[width=84mm]{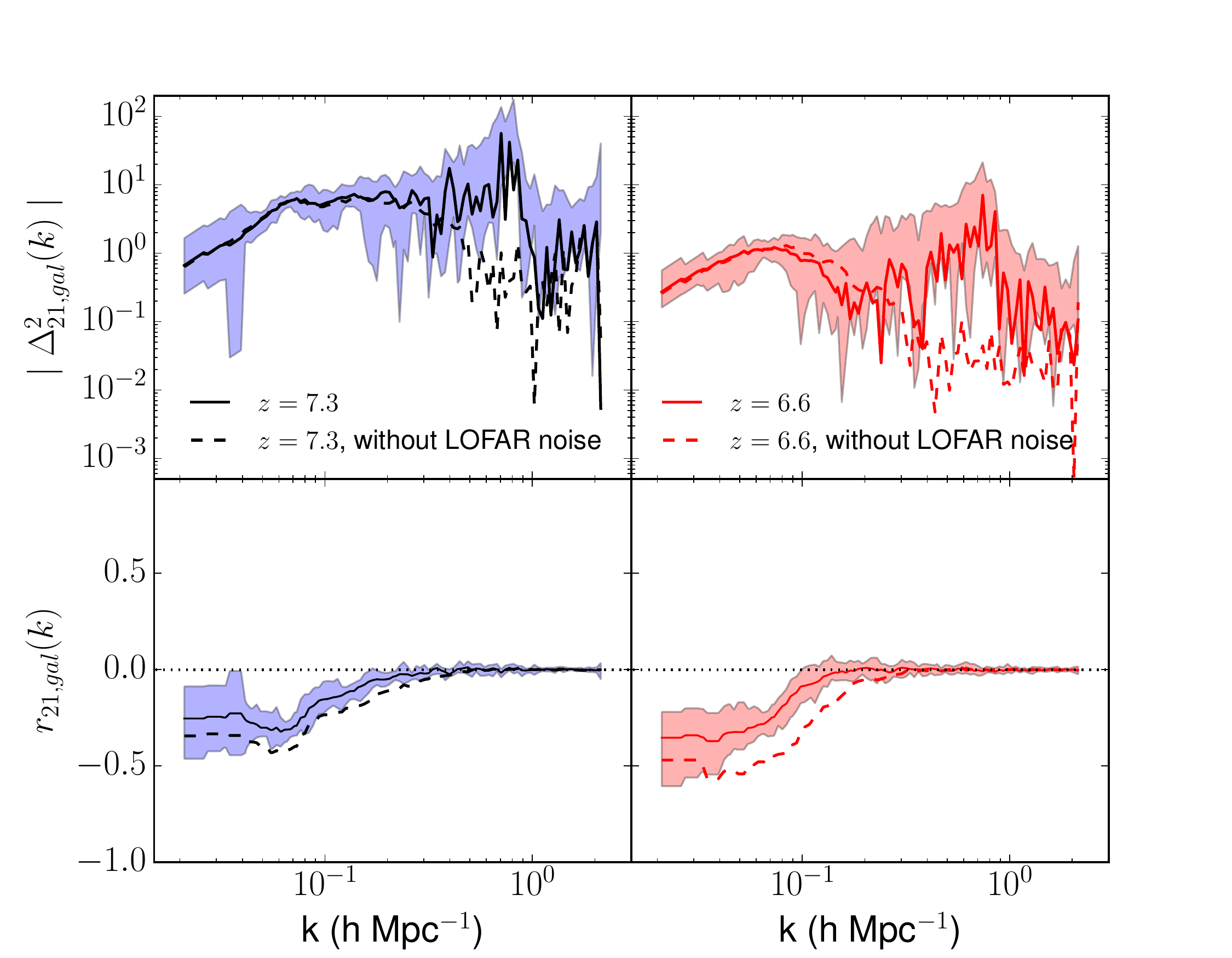}
\caption{Same as Figure~\ref{21gal_lofar}, but for a HSC field of view of 16 deg$^2$ at both redshifts.}
\label{wishful}
\end{figure}

\section{Cross-correlation function}

In this section we present our calculations of the theoretical and observational cross-correlation functions.

\subsection{Theoretical 21cm-galaxy cross-correlation function}

In addition to the cross-power spectrum, we have also computed the cross-correlation function, which shows how the correlation between two fields changes in real space. The cross-correlation function between fields $a$ and $b$ is defined as
$\xi_{a,b}(\mathbf{r})=\langle\delta_a(\mathbf{x})\delta_b(\mathbf{x}+\mathbf{r})\rangle$,
where $\delta(\mathbf{x})$ is the fractional fluctuation of the field at location $\mathbf{x}$. 

The 3D 21cm-galaxy cross-correlation function can then be calculated from the cross-power spectrum as \citep{Park2014}\footnote{Note that when computing the cross-correlation function from the cross-power spectrum, uncertainties arise because of the integration over a finite box size and finite resolution (e.g. uncertainties in the information about the turn over scale; \citealt{Park2014}). However, because of the large box size and number of galaxies, it is computationally much more efficient to compute the 3D 21cm-galaxy cross-correlation function from the cross-power spectrum than directly.}:     
\begin{equation}
\xi_{21,gal}(r)=\frac{1}{(2\pi)^3}\int P_{21,gal}(k)\frac{\sin kr}{kr}4\pi k^2dk.
\end{equation}

In Figure \ref{fig:magnif730} we show the 21cm-galaxy cross-correlation function at $z=7.3$, together with the different terms that contribute to it. $\xi_{\rho,gal}$ shows positive correlation on small scales, where there is an overdensity of both gas and galaxies, and no correlation on large scales. The neutral and galaxy fields, $\xi_{x_{HI},gal}$, are anti-correlated on small scales (where the gas is mostly ionized and there is an overdensity of galaxies), mildly correlated on scales just larger than the typical scale of ionized bubbles (where neutral hydrogen is more abundant), and show no correlation on large scales (where most of the neutral hydrogen resides, but there is a paucity of galaxies). $\xi_{x_{HI}\rho,gal}$ and $\xi_{21,gal}$ behave similarly to $\xi_{x_{HI},gal}$, although $\xi_{x_{HI}\rho,gal}$  turns over to positive values and no correlation on much smaller scales. We can see that the typical scale of ionized regions is $\sim$ 50 $h^{-1}$ Mpc.

\begin{figure}
\centering
\includegraphics[width=84mm]{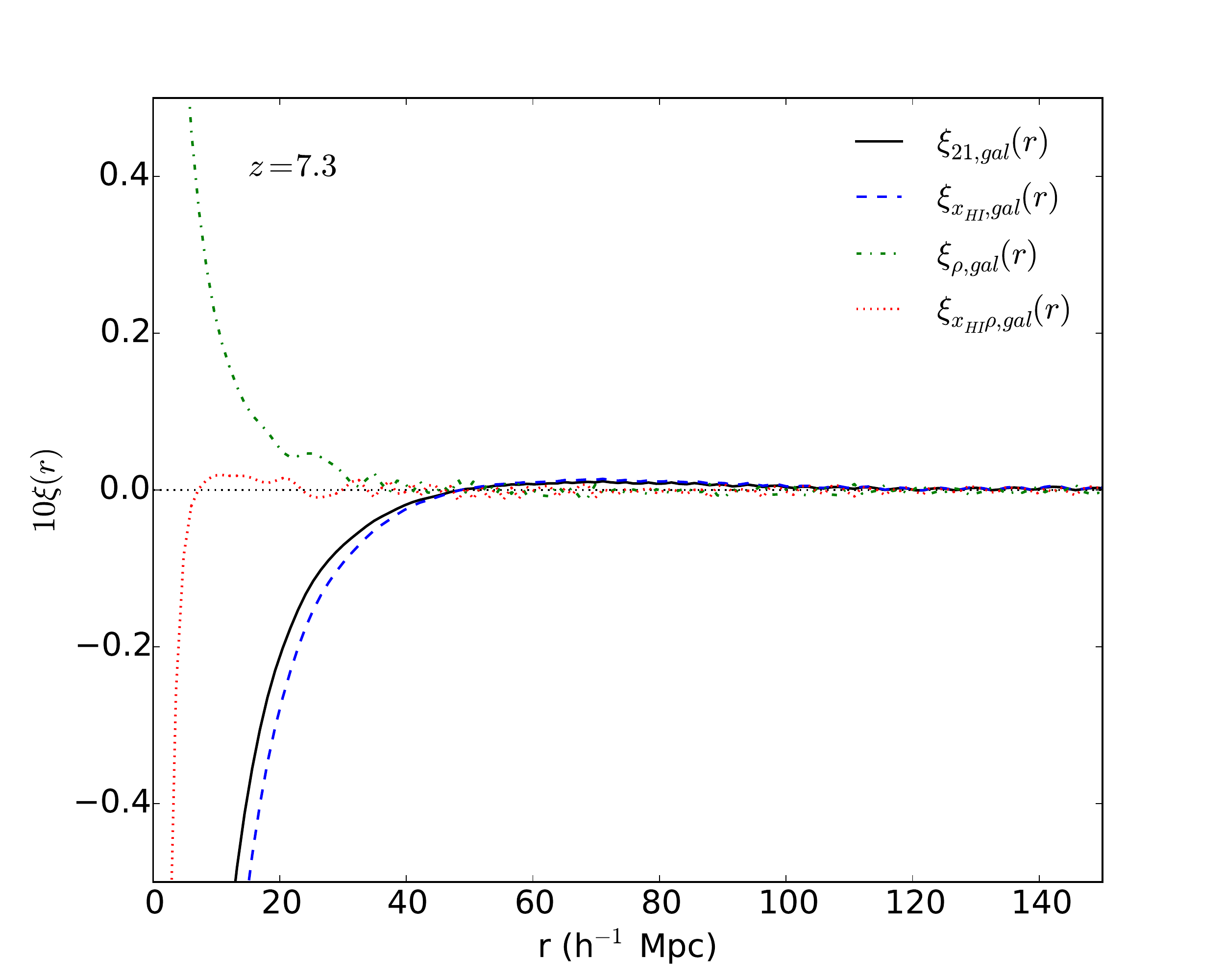}
\caption{Theoretical 3D cross-correlation functions at $z=7.3$ multiplied by 10 for better resolution: 21cm-galaxy, $\xi_{21,gal}$ (black solid line), neutral fraction-galaxy, $\xi_{x_{HI},gal}$ (blue dashed), density-galaxy,  $\xi_{\rho,gal}$ (green dashed-dotted) and neutral density-galaxy, $\xi_{x_{HI}\rho,gal}$ (red dotted). The black dotted line indicates zero correlation. }
\label{fig:magnif730}
\end{figure}

In Figure \ref{fig:cfunc_theo} we show the theoretical 3D 21cm-galaxy cross-correlation function at $z=6.68$, 7.06 and 7.3. The qualitative behaviour of the curves is similar, with an anti-correlation on small scales, indicating the typical scale of the ionized regions, followed by a small positive correlation, and no correlation on larger scales. As for the case of the power spectrum, the anti-correlation is smaller with decreasing redshift due to the fainter 21cm signal. 

\begin{figure}
\centering
\includegraphics[width=84mm]{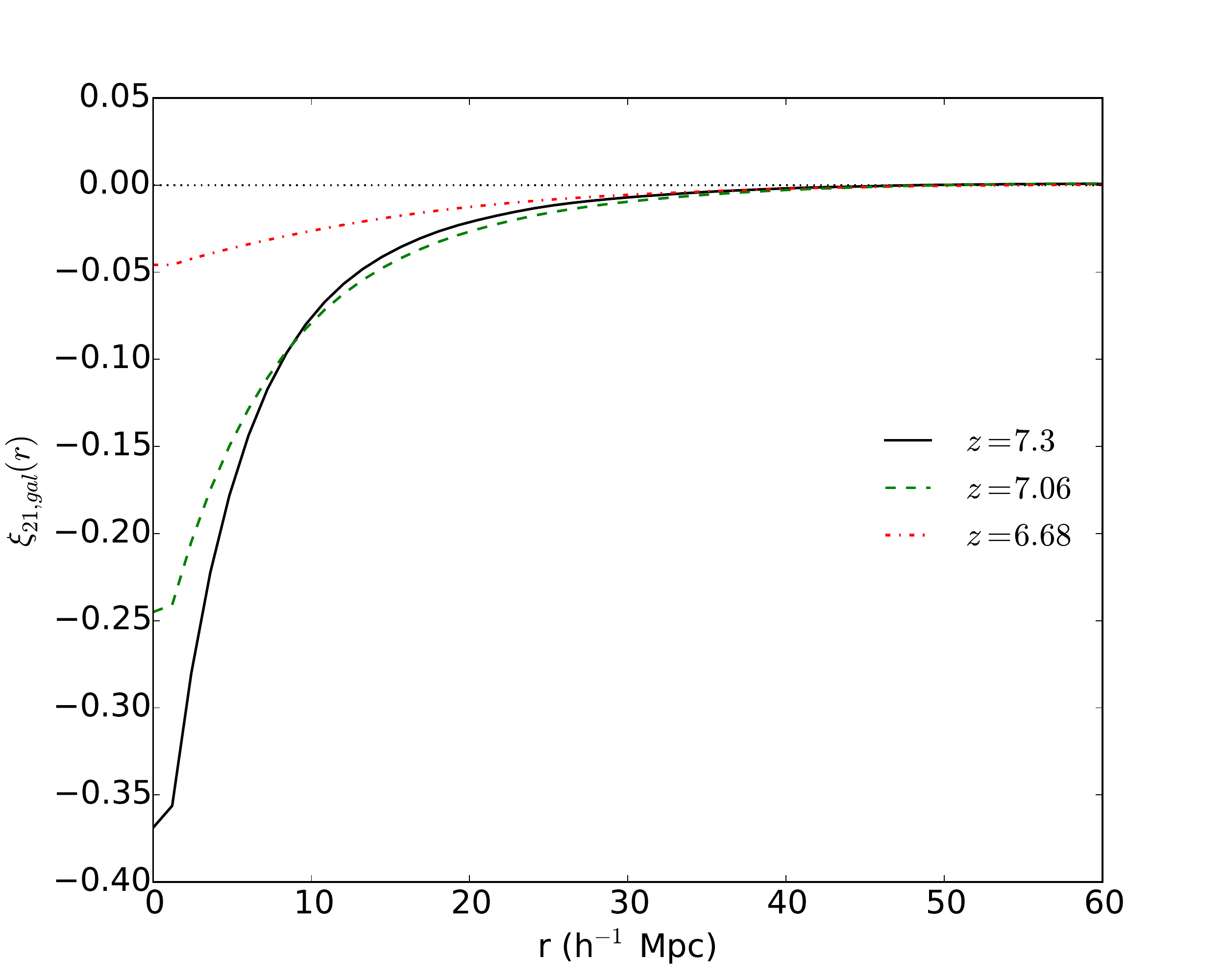}
\caption{Theoretical 3D 21cm-galaxy cross-correlation function at $z=6.68$ (red dashed-dotted line), 7.06 (green dashed) and 7.3 (black). The black dotted line indicates zero correlation. }
\label{fig:cfunc_theo}
\end{figure}

\subsection{Observed 21cm-LAE cross-correlation function}

The observed 2D cross-correlation function can be calculated as:
\begin{equation}
\xi_{21,gal}(r)=\frac{\sum_{\mathbf{x}}\delta_{LAE}(\mathbf{x})\delta_{21}(\mathbf{x}+\mathbf{r})}{N_{pair}(r)},
\end{equation}
where $\delta_{LAE}$ and $\delta_{21}$ are fractional fluctuations of the LAE and 21cm fields, respectively, and $N_{pair}(r)$ is the number of 21cm-LAE pairs at a separation $r$. 

In Figure \ref{fig:cfunc_obs} we plot the 2D 21cm-LAE cross-correlation functions for our reference mock observations (i.e. the equivalent of Fig.~\ref{21gal_lofar}), together with those expected for a larger field of view of 16 deg$^2$ (i.e. the equivalent of Fig.~\ref{wishful}). The observed cross-correlation functions show a behavior similar to the theoretical ones. Noise is large at all scales, resulting in a large scatter. While the average of 10 mock observations for both redshifts shows clear anti-correlation at small scales which goes towards no correlation at large scales, scatter is large, so the detection of the anti-correlation might not be  possible in a single mock observation. The anti-correlations become much clearer in larger fields of view, especially at redshift 6.6. 

\begin{figure}
\centering
\includegraphics[width=84mm]{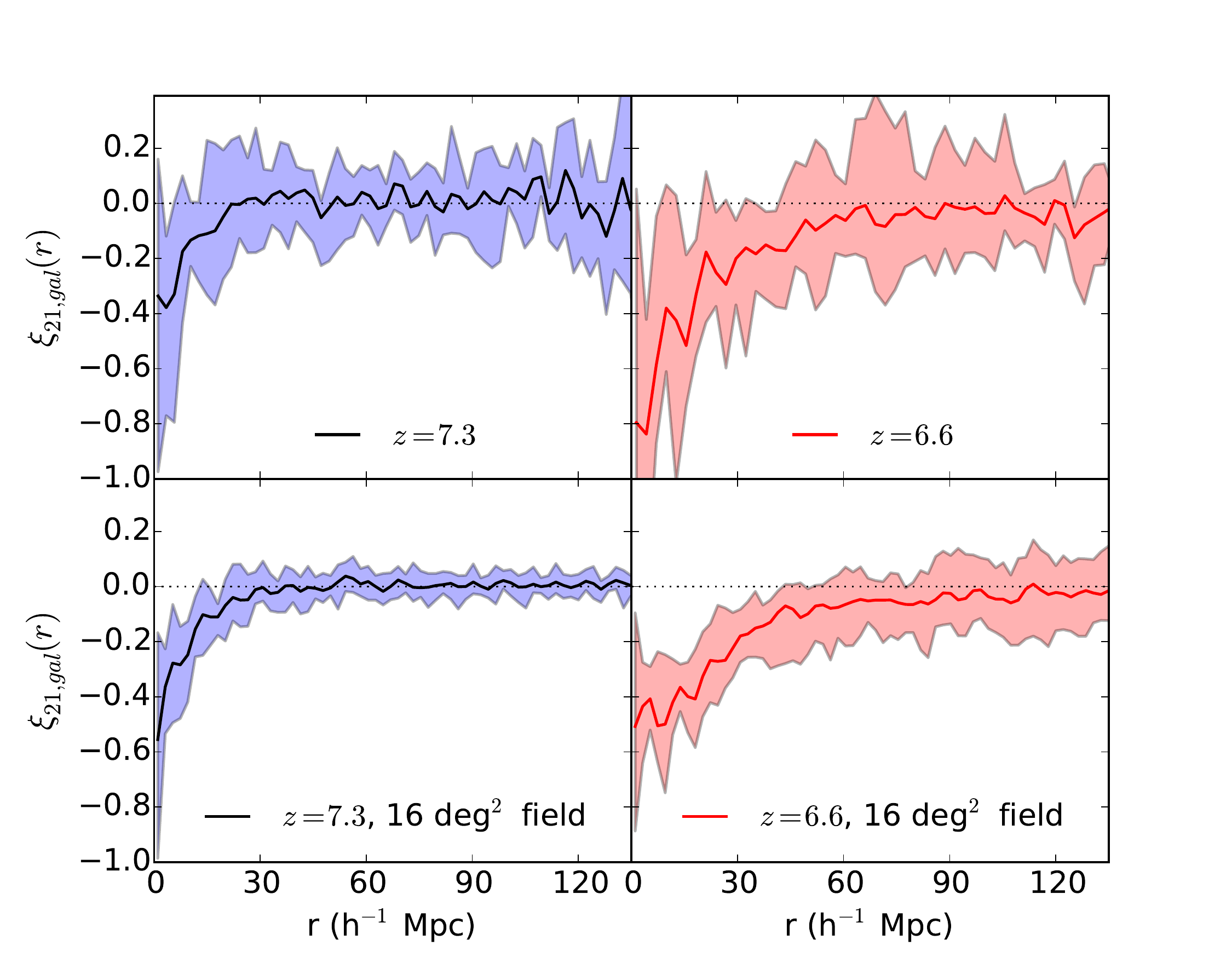}
\caption{2D 21cm-LAE cross-correlation function for our reference mock observations at $z=7.3$ (top left panel) and 6.6 (top right) and for mock observations with fields of view of 16 deg$^2$ at $z=7.3$ (bottom left) and $z=6.6$ (bottom right). The black dotted lines indicate zero correlation and shaded areas indicate scatter from 10 mock observations. }
\label{fig:cfunc_obs}
\end{figure}

\section{Discussion}
\label{sec:discuss}

Observations of 21cm emission and high-$z$ LAEs are extremely challenging, and both will suffer from severe noise problems. Even assuming that foregrounds subtraction will work perfectly, the system noise will still largely exceed the expected signal, in particular at the smaller scales, so that possibly only scales larger than $\sim$ 60 $h^{-1}$ Mpc  (corresponding to $k \sim 0.1 h$~Mpc$^{-1}$) will be accessible by a telescope like LOFAR. In addition, the field of view of HSC is much smaller than that of LOFAR, so that only a fraction of the large scales observed by LOFAR will be covered also by HSC. 

To illustrate this issue further, in Figure~\ref{fig:lofar} we show the 2D 21cm auto-power spectra with and without the LOFAR noise after 600 hours of observation in fields of view of 7 deg$^2$ at $z=6.6$ and 1.7 deg$^2$ at $z=7.3$, i.e. equivalent to the ones of HSC\footnote{Note that in observations of the 21cm auto-power spectra the expectation value of the noise power spectrum can be subtracted from the measurements. However, in observations of the 21cm-LAE cross-power spectra this is not possible, since instrumental effects from LAE observations are also present and their influence can not be treated separately. The same reasoning applies to LAE observations.}. At both redshifts noise on scales smaller than $\sim 60 h^{-1}$ Mpc is orders of magnitude larger than the expected signal, while it decreases gradually at larger scales. Noise on large scales at $z=6.6$ is somewhat larger than at $z=7.3$. 

\begin{figure}
\centering
\includegraphics[width=84mm]{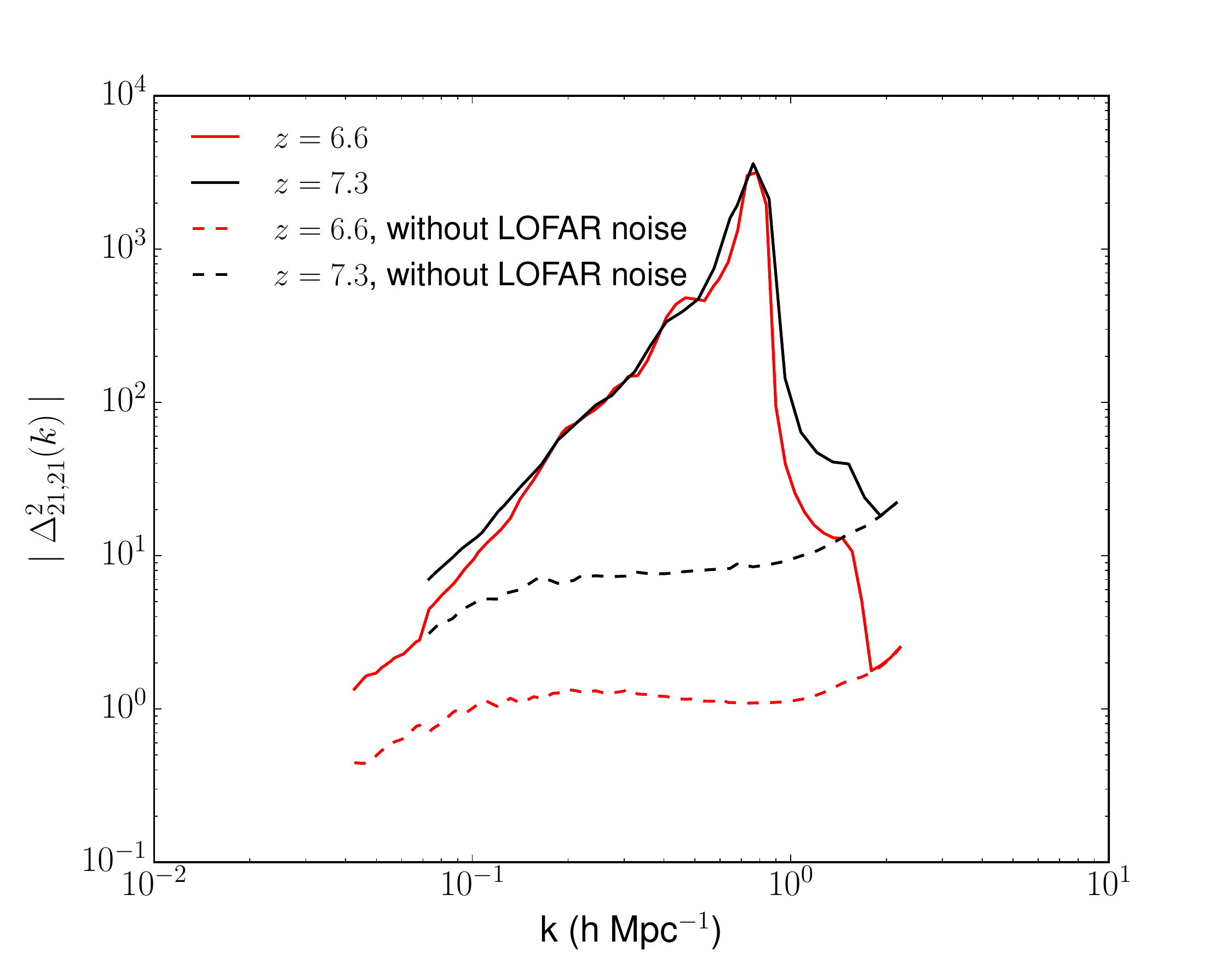}
\caption{2D 21cm auto-power spectrum with (solid lines) and without (dashed) LOFAR noise at $z=6.6$ (red) and 7.3 (black).}
\label{fig:lofar}
\end{figure}

The HSC observations discussed here are groundbreaking, as they will increase the number of detected high-$z$ LAEs by at least one order of magnitude. However, substantial shot noise is still expected, as shown in Figure~\ref{fig:shot_noise}. Observations at both redshifts will be dominated by shot noise at scales below $10 h^{-1}$ Mpc  (i.e. $k>0.6 h$~Mpc$^{-1}$) at $z$=6.6, and $30 h^{-1}$ Mpc  (i.e. $k>0.2 h$~Mpc$^{-1}$) at $z$=7.3, while at large scales the LAEs auto-power spectrum is stronger than that of the shot noise, in particular at $z=6.6$.

\begin{figure}
\centering
\includegraphics[width=84mm]{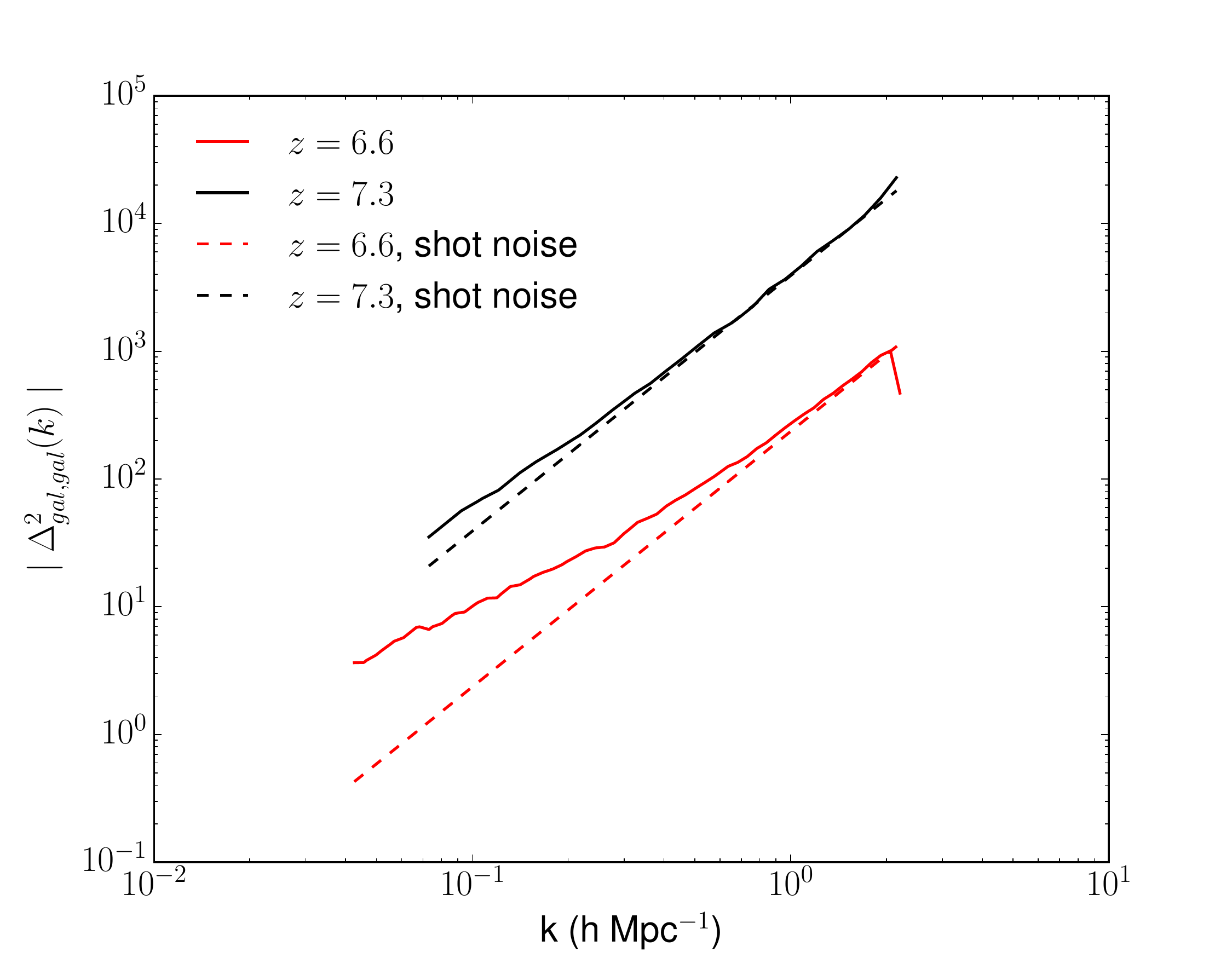}
\caption{2D LAE auto-power spectra (solid lines) and shot noise power spectra (dashed) at $z=6.6$ (red) and 7.3 (black). }
\label{fig:shot_noise}
\end{figure}

Since the 21cm-LAE cross-power spectrum will be affected by noise from both instruments, we expect to be able to probe only scales larger than $\sim 60 h^{-1}$ Mpc  (i.e. $k<0.1 h$~Mpc$^{-1}$). Such scales will still have shot noise, in particular at $z=7.3$, but this should not prevent the detection of an anti-correlation. 

Stronger anti-correlation could be detected by reducing the noise, e.g. increasing the integration time for 21cm observations ($\sigma_{noise}\sim t_{int}^{-1/2}$ for LOFAR; Eq. \ref{eq:noise}), or with a larger field of view. The latter would increase the number of observed LAEs and thus reduce the shot noise and extend the number of observed $k$-modes.

In 21cm-LAE cross-correlation function noise is large at all scales. This is because, unlike in cross-power spectrum, noise does not get separated by its $k$-modes, and thus it is equally distributed on all scales. Comparing cross-correlation functions at $z=7.3$ for our reference mock observations and for mock observations with a larger field of view (Fig. \ref{fig:cfunc_obs}), in the latter case a larger amplitude of the anti-correlation as well as a smaller scatter can be observed because of a reduction of the noise component. While the scatter is smaller also at $z=6.6$, the amplitude is not increased. We suggest that this is due to the LOFAR noise, which has a stronger effect at $z=6.6$, despite being smaller in absolute terms. The noise should also be responsible for the positive correlation observed at large scales in the top panels of Figures~\ref{fig:660smooth} and \ref{fig:730smooth}.

To reduce the noise levels we also smoothed the 21cm field with a Gaussian filter with standard deviations of $\sigma=2$, 5 and 10 (with smoothing radii $4.14\ h^{-1}$ Mpc, $10.35\ h^{-1}$ Mpc and $20.70\ h^{-1}$ Mpc, respectively) at $z=6.6$ (Fig. \ref{fig:660smooth}) and $\sigma=1$, 2 and 5 (with smoothing radii $2.07 h^{-1}$ Mpc, $4.14\ h^{-1}$ Mpc and $10.35\ h^{-1}$ Mpc, respectively) at $z=7.3$ (Fig. \ref{fig:730smooth}). We clearly see that both the average cross-correlation function and the scatter become smoother with increasing $\sigma$. However, by smoothing the field we also loose information (e.g. in terms of anti-correlation amplitude), which is visible when comparing results with different $\sigma$. Smoothing the signal at $z=6.6$ reduces the noise on small scales enough that the anti-correlation becomes clear even for the largest scatter. At $z=7.3$, instead, the shot noise is larger because of the smaller LAE sample, so that even after smoothing the scatter on small scales remains large.

While smoothing reduces noise in the cross-correlation function (which remains though still noisy on all scales), it is not helpful when applied to the 21cm-galaxy cross-power spectrum. The reason for this is that because the noise is being separated by its $k$-modes, smoothing would affect only the small scales which would still be over-contaminated by shot noise. However, the separation of noise by its $k$-modes is exactly what makes large scales observable and the 21cm-galaxy cross-power spectrum a more useful probe of reionization.

\begin{figure}
\centering
\includegraphics[width=84mm]{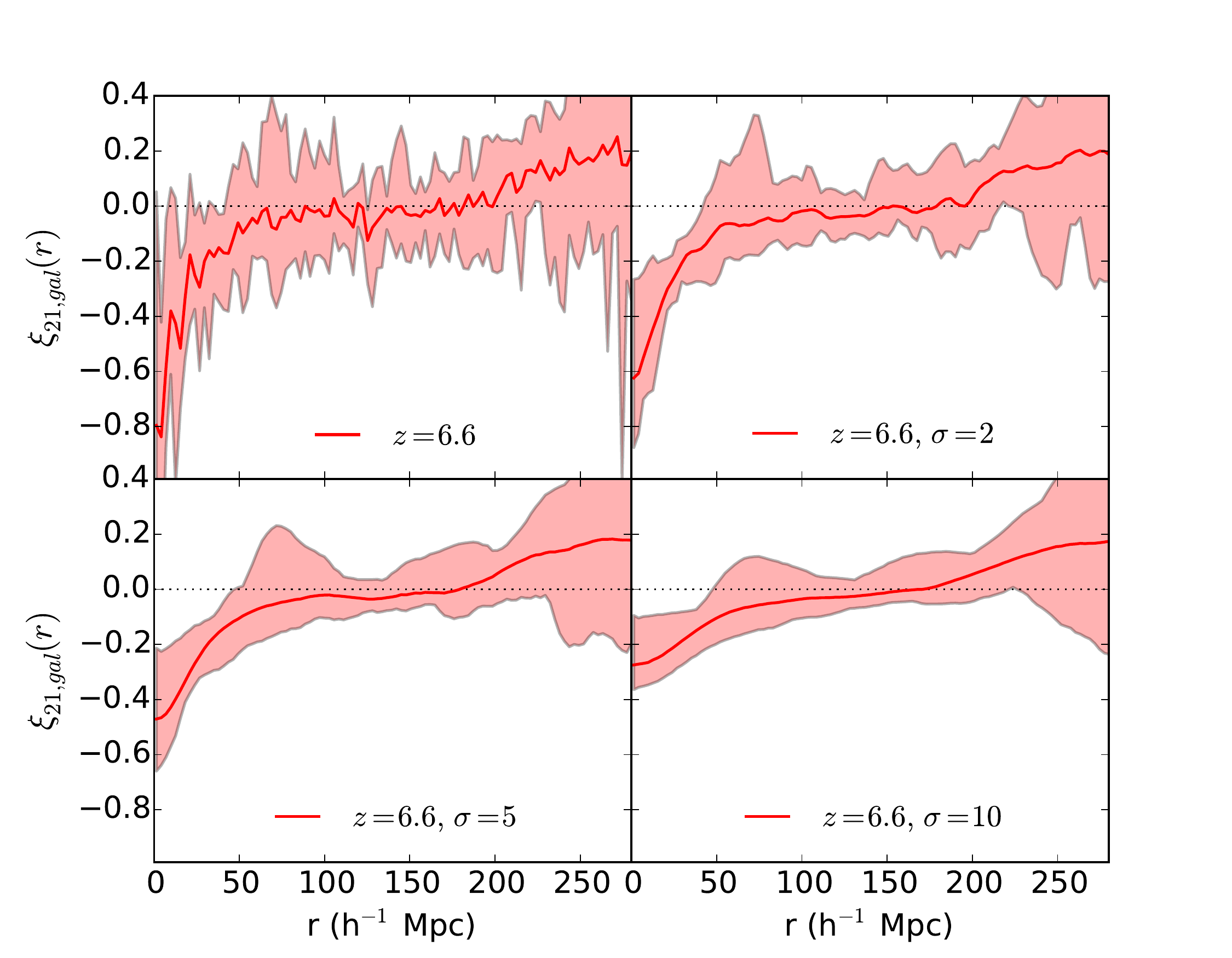}
\caption{2D 21cm-LAE cross-correlation function at $z=6.6$ (top left panel) using a Gaussian filter with standard deviation $\sigma=2$ (top right), $\sigma=5$ (bottom left) and $\sigma=10$ (bottom right). The black dotted lines indicate zero correlation and shaded areas scatter from 10 mock observations. }
\label{fig:660smooth}
\end{figure}

\begin{figure}
\centering
\includegraphics[width=84mm]{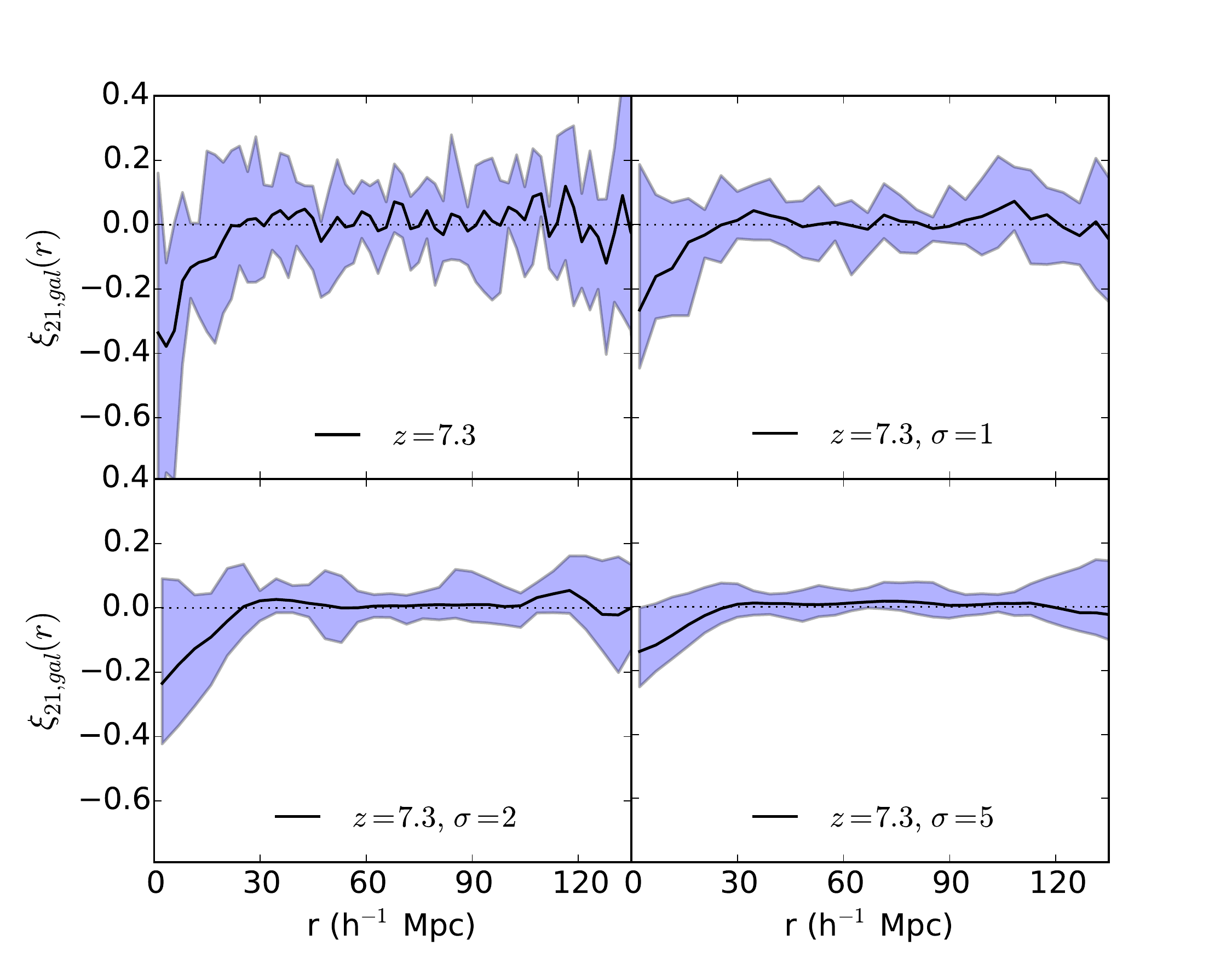}
\caption{2D 21cm-LAE cross-correlation function at $z=7.3$ (top left panel) using a Gaussian filter with standard deviation $\sigma=1$ (top right), $\sigma=2$ (bottom left) and $\sigma=5$ (bottom right). The black dotted lines indicate zero correlation and shaded areas scatter from 10 mock observations. }
\label{fig:730smooth}
\end{figure}

\section{Conclusions}

In this paper we have investigated the properties of the 21cm-galaxy cross-power spectrum and its observability with upcoming observations by LOFAR and HSC. To this aim, we have used snapshots at redshifts 6.68, 7.06 and 7.3 from N-body + radiative transfer simulations \citep{Iliev2014}, processed with a  Ly$\alpha$ radiative transfer code \citep{Jensen2013}. Our theoretical 3D 21cm-galaxy cross-power spectrum agrees with previous investigations (i.e. \citealt{Lidz2009,W2013}). More specifically, we are able to recover the same redshift dependence and shape, with a distinct turnover point indicating the typical scale of ionized bubbles. We confirm that the 21cm-galaxy cross-power spectrum could provide information on the progress of reionization and the typical size of HII regions at different redshifts.

The measured 21cm-LAE cross-power spectrum suffers from projection effects (as it is 2D), as well as from noise in both radio and LAEs detections.
LOFAR recently started observations of 21cm emission from neutral hydrogen in the redshift range $z=6-11.4$ \citep{Yatawatta2013, Jelic2014}, while HSC will also soon start its observational campaign with two narrow-band filters searching for LAEs at $z=6.6$ and 7.3 (M. Ouchi 2012, private conversation). Both telescopes plan to observe the ELAIS-N1 field at $z=6.6$, making it possible to detect the 21cm-galaxy cross-power spectrum.  
We constructed mock observations specifically tailored to match LOFAR and HSC campaigns at redshifts 6.6 and 7.3. Our mock observations show that despite the observed spectra being much noisier than the corresponding theoretical 3D ones, dependence of the normalization on redshift (i.e. amount of HI) is clearly visible, as well as the anti-correlation between the two fields, with a cross-correlation coefficient $r_{21,gal}\approx-0.20$  at levels of significance of $p=0.04$ at $z=6.6$ and $p=0.048$ at $z=7.3$. However, the turnover point can not be clearly determined because small scales will be overwhelmed by noise.

We also investigated properties and observability of the 21cm-galaxy cross-correlation functions, which are expected to be negative on small scales, mildly correlated on scales just larger than the typical size of ionized regions, and show no correlation on even larger scales. This agrees well with predictions of the 21cm-galaxy cross-correlation function by \citet{Park2014}. Despite observational effects like noise and galaxy number densities, observed correlation functions should retain the theoretical shape. However, unlike the observed 21cm-LAE cross-power spectrum, the correlation function suffers from a strong noise on all scales (as it does not get separated by its $k$-modes), thus uncertainties in the signal will be large.

In summary, the 21cm-LAE cross-power spectrum is a powerful probe of the EoR which could provide invaluable information on the progress of reionization and the typical scale of ionized regions at different redshifts. Observations with LOFAR and HSC will finally make detection of the 21cm-LAE cross-power spectrum possible at redshift 6.6, as they both plan to observe the ELAIS-N1 field. These observations are going to be very challenging and have substantial problems with noise, but they will still be able to detect the large scales of the cross-power spectrum, which is expected to show an anti-correlation between the two fields. 

\section*{Acknowledgements}
VJ would like to thank the Netherlands Foundation for Scientific Research (NWO) for financial support through VENI grant 639.041.336. ITI was supported by the Science and Technology Facilities Council [grant numbers ST/F002858/1, ST/I000976/1 and ST/L000652/1]. GM is supported by Swedish Research Council grant 2012-4144. LVEK and AG acknowledge the financial support from the European Research Council under ERC-Starting Grant FIRSTLIGHT - 258942. The authors would also like to thank Gianni Bernardi for useful insights and comments.

The authors acknowledge Paul Shapiro for permission to use the simulations on which this paper was based, described in Iliev et al. (2014). That work was supported in part by grants and allocations of which Shapiro is the P.I., including U.S. NSF grant AST-1009799, NASA grant
NNX11AE09G, NASA/JPL grant RSA Nos. 1492788 and 1515294, and supercomputer resources from NSF XSEDE grant TG-AST090005 and the Texas Advanced Computing Center (TACC) at the University of Texas at Austin.
Some of the numerical computations were done on the Apollo cluster at The University of Sussex and the Sciama High Performance Compute (HPC) cluster which is supported by the ICG, SEPNet and the University of Portsmouth. Part of the computations were performed on the GPC supercomputer at the SciNet HPC Consortium (courtesy Ue-Li Pen). SciNet is funded by: the Canada Foundation for Innovation under the auspices of Compute Canada; the Government of Ontario; Ontario Research Fund - Research Excellence; and the University of Toronto. The authors thank Kyungjin Ahn for providing the recipe for including sub-resolution sources in the simulation volume.

\bibliography{21gal_ref1}
\bsp

\label{lastpage}

\end{document}